\documentclass[11pt]{article}
\usepackage{graphics,epsfig}
\usepackage[]{graphicx}
\usepackage{latexsym}
\usepackage{amsmath, amsthm, amsfonts, amssymb}
\usepackage{verbatim}

\begin{document}

\newtheorem{theo}{Theorem}[section]
\newtheorem{definition}[theo]{Definition}
\newtheorem{lem}[theo]{Lemma}
\newtheorem{prop}[theo]{Proposition}
\newtheorem{coro}[theo]{Corollary}
\newtheorem{exam}[theo]{Example}
\newtheorem{rema}[theo]{Remark}
\newtheorem{example}[theo]{Example}
\newcommand{\ninv}{\mathord{\sim}} 
\newtheorem{axiom}[theo]{Axiom}

\title{A Quantum Logical and Geometrical Approach to the Study of Improper Mixtures}

\author{{\sc Graciela
Domenech}\thanks{%
Fellow of the Consejo Nacional de Investigaciones Cient\'{\i}ficas y
T\'ecnicas (CONICET)}$^{,\ 1}$  {\sc ,} \ {\sc Federico Holik}$^{1}$
\ {\sc and} \ {\sc C\'{e}sar Massri}$^{2}$}

\maketitle

\begin{center}

\begin{small}
1- Instituto de Astronom\'{\i}a y F\'{\i}sica del Espacio (IAFE)\\
Casilla de Correo 67, Sucursal 28, 1428 - Buenos Aires, Argentina\\
2- Departamento de Matem\'{a}tica - Facultad de Ciencias Exactas y
Naturales\\ Universidad de Buenos Aires - Pabell\'{o}n I, Ciudad
Universitaria \\ Buenos Aires, Argentina

\end{small}
\end{center}

\vspace{1cm}

\begin{abstract}
\noindent We study improper mixtures from a quantum logical and
geometrical point of view. Taking into account the fact that
improper mixtures do not admit an ignorance interpretation and must
be considered as states in their own right, we do not follow the
standard approach which considers improper mixtures as measures over
the algebra of projections. Instead of it, we use the convex set of
states in order to construct a new lattice whose atoms are all
physical states: pure states and improper mixtures. This is done in
order to overcome one of the problems which appear in the standard
quantum logical formalism, namely, that for a subsystem of a larger
system in an entangled state, the conjunction of all actual
properties of the subsystem does not yield its actual state. In
fact, its state is an improper mixture and cannot be represented in
the von Neumann lattice as a minimal property which determines all
other properties as is the case for pure states or classical
systems. The new lattice also contains all propositions of the von
Neumann lattice. We argue that this extension expresses in an
algebraic form the fact that -alike the classical case- quantum
interactions produce non trivial correlations between the systems.
Finally, we study the maps which can be defined between the extended
lattice of a compound system and the lattices of its subsystems.
\end{abstract}
\bigskip
\noindent

\begin{small}
\centerline{\em Key words: quantum logic, convex set of states,
entanglement.}
\end{small}

\bibliography{pom}

\section{Introduction}

Non-separability of the states of quantum systems is considered with
continuously growing interest in relation to quantum information
theory. In fact, today entanglement is regarded not only as a
feature that gives rise to interesting foundational questions. It is
considered also as a powerful resource for quantum information
processing. In this paper we pose the problem of studying
non-separability with algebraic and geometrical tools related to
quantum logic (QL).

The algebraic approach to the formalization of quantum mechanics was
initiated by Birkhoff and von Neumann \cite{BvN}, who gave it the
name of ``quantum logic''. Although an algebraic structure, for
historical reasons it has conserved its name. QL was developed
mainly by Mackey \cite{mackey57}, Jauch \cite{jauch}, Piron
\cite{piron}, Kalmbach \cite{kalm83, kalm86}, Varadarajan
\cite{vadar68, vadar70}, Greechie \cite{greechie81}, Gudder
\cite{gudderlibro78},  Giuntini \cite{giunt91}, Pt\'{a}k and
Pulmannova \cite{pp91}, Beltrametti and Cassinelli \cite{belcas81},
among others. For a complete bibliography see for example
\cite{dallachiaragiuntinilibro} and \cite{dvupulmlibro}. The Geneva
school of QL extended this research to analysis of compound systems.
The first results where obtained  by Aerts and Daubechies
\cite{aertsdaub1, aertsdaub2} and Randall and Foulis \cite{FR81}.

In the tradition of the quantum logical research, a property of (or
a proposition about) a quantum system is related to a closed
subspace of the Hilbert space ${\mathcal H}$ of its (pure) states
or, analogously, to the projector operator onto that subspace.
Moreover, each projector is associated to a dichotomic question
about the actuality of the property {\rm \cite[pg. 247]{vN}}. A
physical magnitude ${\mathcal M}$ is represented by an operator $M$
acting over the state space. For bounded self-adjoint operators,
conditions for the existence of the spectral decomposition
$M=\sum_{i} a_i P_i=\sum_{i} a_i |a_i\rangle\langle a_i|$ are
satisfied (along this work we will restrict the study to the finite
dimensional case). The real numbers $a_i$ are related to the
outcomes of measurements of the magnitude ${\mathcal M}$ and
projectors $|a_i\rangle\langle a_i|$ to the mentioned properties.
The physical properties of the system are organized in the lattice
of closed subspaces ${\mathcal L}({\mathcal H})$ that, for the
finite dimensional case, is a modular lattice \cite{MM}. In this
frame, the pure state of the system is represented by the meet (i.e.
the lattice infimum) of all actual properties, more on this below. A
comprehensive description of QL in present terminology may be found
in \cite{svozillibro}.

Mixed states represented by density operators had a secondary role
in the classical treatise by von Neumann because they did not add
new conceptual features to pure states. In fact, in his book,
mixtures meant ``statistical mixtures'' of pure states {\rm
\cite[pg. 328]{vN}}, which are known in the literature as ``proper
mixtures'' {\rm \cite[Ch. 6]{d'esp}}. They usually represent the
states of realistic physical systems whose preparation is not well
described by pure states.

Today we know that the restriction to pure states and their mixtures
is unduly because there are also ``improper mixtures'' and they do
not admit an ignorance interpretation (\cite{d'esp},
\cite{Mittelstaedt}, \cite{Kirkpatrik}, \cite{ReplytoKirkpatrik},
\cite{quaternionicos}, \cite{Ufano}). This fact is an expression of
one of the main features of quantum systems, namely
non-separability. Improper mixtures are now considered as states on
their own right, and they appear for example, in processes like
measurements on some degrees of freedom of the system, and also when
considering one system in a set of interacting systems. In fact, in
each (non trivial) case in which a part of the system is considered,
we have to deal with improper mixtures. (Also for statistical
mixtures the ignorance interpretation becomes untenable in cases of
nonunique decomposability of the density operator \cite[Ch.
2]{belcas81}.)

In the standard formulation of QL, mixtures as well as pure states
are included as measures over the lattice of projections \cite[Ch.
3]{mikloredeilibro}, that is, a state $s$ is a function:
$$s:\mathcal{L}(\mathcal{H})\longrightarrow [0;1]$$
\noindent such that:
\begin{enumerate}
\item $s(\textbf{0})=0$ ($\textbf{0}$ is the null subspace).
\item For any pairwise orthogonal family of projections ${P_{j}}$, $s(\sum_{j}P_{j})=\sum_{j}s(P_{j})$
\end{enumerate}
In a similar way, in classical mechanics statistical distributions
are represented as measures over the phase space. But while pure
states can be put in a bijective correspondence to the atoms of
$\mathcal{L}(\mathcal{H})$, this is not the case for mixtures of
neither kind. On the contrary, the standard formulation of $QL$
treats improper mixtures in an analogous way as classical
statistical distributions. But improper mixtures have a very
different physical content, because they do not admit an ignorance
interpretation. After a brief review of the problem of quantum
non-separability in Section \ref{s:Quantum Non-Separability}, we
turn in Section  \ref{s:explicamosporque} to the reasons why this
difference leads to a dead end when compound systems are considered
from the standard quantum logical point of view. We also discuss
that the physical necessity to consider mixtures indicates that the
algebraic structure of the properties of compound systems should be
studied in a frame that takes into account the fact that density
operators are states in their own right. We show in Section
\ref{s:New language} that a frame with these characteristics can be
built by enlarging the scope of standard QL. We do this by
constructing a lattice based on the convex set of density operators
which incorporates improper mixtures as atoms of the lattice. Then,
in Section \ref{s:The Relationship} we study the relationship
between this lattice and the lattices of its subsystems and show how
our construction overcomes the problem posed in Section
\ref{s:explicamosporque}. Finally, we draw our conclusions in
Section \ref{s:Conclusions}.

\section{Quantum non-separability}\label{s:Quantum Non-Separability}

We briefly review here the main arguments and results of the
analysis of non-separability and relate them to the frame of quantum
logical research for the sake of completeness. We start by analyzing
classical compound systems in order to illustrate their differences
with the quantum case.

\subsection{Classical systems}

When considering in classical mechanics two systems $S_{1}$ and
$S_{2}$ and their own state spaces $\Gamma_{1}$ and $\Gamma_{2}$
(or, analogously, two parts of a single system), the state space
$\Gamma$ of the composite system is the cartesian product $\Gamma=
\Gamma_{1}\times\Gamma_{2}$ of the phase spaces of the individual
systems, independently of the kind of interaction between both of
them. The physical intuition behind this fact is that, no matter how
they interact, every interesting magnitude corresponding to the
parts and the whole may be written in terms of the points in phase
space.

In the logical approach, classical properties are associated with
subsets of the phase space, precisely with the subsets consisting of
the points corresponding to those states such that, when being in
them, one may say that the system has the mentioned property. Thus,
subsets of $\Gamma$ are good representatives of the properties of a
classical system. The power set $\wp(\Gamma)$  of $\Gamma$,
partially ordered by set inclusion $\subseteq$ (the implication) and
equipped with set intersection $\cap$ as the meet operation, set
union $\cup$ as the join operation and relative complement $'$ as
the complement operation gives rise to a complete Boolean lattice
$<\wp(\Gamma),\ \cap,\ \cup,\ ',\ \mathbf{0},\ \mathbf{1}>$ where
$\mathbf{0}$ is the empty set $\emptyset$ and $\mathbf{1}$ is the
total space $\Gamma$. According to the standard interpretation,
partial order and lattice operations may be put in correspondence
with the connectives $and$, $or$ $not$ and the $material\
implication$ of classical logic.

In this frame, the points $(p,q)\in\Gamma$ (pure states of a
classical system) represent pieces of information that are maximal
and logically complete. They are maximal because they represent the
maximum of information about the system that cannot be consistently
extended (any desired magnitude is a function of $(p,\ q)$) and
complete in the sense that they semantically decide any property
\cite{dallachiaragiuntinilibro}. Statistical mixtures are
represented by measurable functions:
$$\sigma:\Gamma\longrightarrow [0;1]$$\label{classical statistical
mixture} \noindent such that
$$\int_{\Gamma}\sigma(p,q)d^{3}pd^{3}q=1$$

We point out that statistical mixtures are not fundamental objects
in classical mechanics, in the sense that they admit an ignorance
interpretation. They appear as a state of affairs in which the
observer cannot access to an information which lies objectively in
the system. Although the physical status of quantum improper
mixtures is very different, they are treated in a similar way as
classical mixtures by standard QL. We discuss in Section
\ref{s:explicamosporque} how this misleading treatment leads to
problems.

When considering two systems, it is meaningful to organize the whole
set of their properties in the corresponding (Boolean) lattice built
up as the cartesian product of the individual lattices. Informally
one may say that each factor lattice corresponds to the properties
of each physical system. More precisely, in the category of lattices
as objects and lattice morphisms as arrows, the cartesian product of
lattices is the categorial product. This category is $Ens$, and the
cartesian product is the categorial product in $Ens$.

\subsection{Quantum systems}

The quantum case is completely different. When two or more systems
are considered together, the state space of their pure states is
taken to be the tensor product of their Hilbert spaces. Given the
Hilbert state spaces $\mathcal{H}_{1}$ and $\mathcal{H}_{2}$ as
representatives of two systems, the pure states of the compound
system are given by rays in the tensor product space
$\mathcal{H}=\mathcal{H}_{1}\otimes\mathcal{H}_{2}$. But it is not
true --as a naive classical analogy would suggest-- that any pure
state of the compound system factorizes after the interaction in
pure states of the subsystems, and that they evolve with their own
Hamiltonian operators \cite{Mittelstaedt, aertsjmp83}. The
mathematics behind the persistence of entanglement is the lack of a
product of lattices and even posets \cite{aertsQL81,
aertsrepmathphys84, dvu95}. A product of structures is available for
weaker structures \cite[Ch. 4]{dvupulmlibro} but those structures,
though mathematically very valuable and promising, have a less
direct relation with the standard formalism of quantum mechanics.

In the standard quantum logical approach, properties (or
propositions regarding the quantum system) are in correspondence
with closed subspaces of Hilbert space $\mathcal{H}$. The set of
subspaces ${\mathcal{C}}({\mathcal{H}})$ with the partial order
defined by set inclusion $\subseteq$, intersection of subspaces
$\cap$ as the lattice meet, closed linear spam of subspaces $\oplus$
as the lattice join and orthocomplementation $\neg$ as lattice
complement, gives rise in the finite dimensional case to a modular
lattice ${\mathcal{L}}(\mathcal{H})=<{\mathcal{C}}({\mathcal{H}}),\
\cap,\ \oplus,\ \neg,\ \mathbf{0},\ \mathbf{1}>$ where $\mathbf{0}$
is the empty set $\emptyset$ and $\mathbf{1}$ is the total space
$\mathcal{H}$. We will refer to this lattice as
${\mathcal{L}}_{v\mathcal{N}}$, the `von Neumann lattice'.

When trying to repeat the classical procedure of taking the tensor
product of the lattices of the properties of two systems to obtain
the lattice of the properties of the composite the procedure fails
\cite{aertsjmp84,FR81}. Mathematically, this is the expression of
the fact that the category of Hilbert lattices as objects and
lattice morphisms as arrows has not a categorial product because of
the failure of orthocomplementation. This problem is studied for
example in \cite{aertsdaub2,gudderlibro78}. Attempts to vary the
conditions that define the product of lattices have been made
\cite{pulmJMP85, ischi2005}, but in all cases it results that the
Hilbert lattice factorizes only in the case in which one of the
factors is a Boolean lattice or when systems have never interacted.
For a complete review, see \cite{dvu95}.

Let us briefly recall the defining properties of the tensor product
of a finite collection of vector spaces in order to discuss the main
features that make the difference with the classical case. Let us
first define (following \cite{bratellilibro})
$\otimes\mathcal{H}_{i}$ as the unique vector space which satisfies
the following properties:

\begin{enumerate}

\item for each family $\{|x_{i}\rangle\}$, $|x_{i}\rangle\in\mathcal{H}_{i}$,
there exists an element $\otimes_{i}
|x_{i}\rangle\in\otimes_{i}\mathcal{H}_{1}$ depending multilinearly
on the $\{|x_{i}\rangle\}$. All vectors in
$\otimes_{i}\mathcal{H}_{i}$ are finite linear combinations of such
elements.

\item (universal property) for each multilinear mapping $\pi$ of the
product of the $\mathcal{H}_{i}$ into a vector space $Y$, there
exists a unique linear map
$\varphi:\otimes_{i}\mathcal{H}_{i}\longrightarrow Y$ such that
$$\varphi(\otimes_{i}
|x_{i}\rangle)=\pi(\{|x_{i}\rangle\})$$ for all
$|x_{i}\rangle\in\mathcal{H}_{i}$.

\item (associativity) for each partition $\cup_{k}I_{k}$ of
$\{1,\cdots,n\}$ there exists a unique isomorphism from
$\otimes_{i}\mathcal{H}_{i}$ onto $\otimes_{k}(\otimes_{i\in
I_{k}}\mathcal{H}_{i})$ transforming $\otimes_{i}|x_{i}\rangle$ into
$\otimes_{k}(\otimes_{i\in I_{k}}|x_{i}\rangle)$.

\end{enumerate}

\noindent When the spaces $\mathcal{H}_{i}$ are Hilbert spaces, it
is possible to define an inner product on $\otimes\mathcal{H}_{i}$
by extending the following definition by linearity:

$$(\otimes_{i}|x_{i}\rangle,\otimes_{i}|y_{i}\rangle)=\prod_{i}(|x_{i}\rangle,|y_{i}\rangle)$$

\noindent Note that as we are using Dirac notation, we may write
$\langle x_{i}|y_{i}\rangle$ instead of
$(|x_{i}\rangle,|y_{i}\rangle)$. The completion of
$\otimes\mathcal{H}_{i}$ in the associated norm is the tensor
product of the Hilbert spaces $\otimes_{i}\mathcal{H}_{i}$. Thus we
see that the tensor product of Hilbert spaces is in essence a
multilinear extension of the direct product. From a physical point
of view, it is for this reason that the state of the joint system
contains much more information than `the sum' of the information
contained in the states of its parts.

This feature of quantum systems may be regarded as a strange fact
when using classical reasoning, but it not strange at all in a
landscape where the superposition principle holds. Given two systems
$S_{1}$ and $S_{2}$, if we prepare them independently in states
$|a\rangle$ and $|b\rangle$ respectively, then we would have
something like the direct product of the states of both systems
$|a\rangle\times|b\rangle$ for the state of the joint system. We
could perform also different preparations and obtain
$|a'\rangle\times|b'\rangle$. Then, if there are no superselection
rules and according to the superposition principle, it is quite
natural to suppose that it is at least in principle possible to
prepare the superposition state of the form
$\alpha|a\rangle\otimes|b\rangle+\beta|a'\rangle\otimes|b'\rangle$,
and so, we \emph{need} $\otimes$ instead of $\times$. This last
state is not a product of the states of the parties. It is for this
reason that the product in quantum mechanics has to be the
multilinear extension of the direct product.

Let us now briefly review the standard relationship between the
states of the joint system and the states of the subsystems. If
$\{|x_{k}^{(i)}\rangle\}$ is an orthonormal basis for
$\mathcal{H}_{i}$, then

$${\otimes_{i=1}^{n}|x_{k_{i}}^{(i)}\rangle}$$

\noindent forms a basis of $\otimes_{i}\mathcal{H}_{i}$. Let us
focus for simplicity on the case of two systems, $S_{1}$ and
$S_{2}$. If $\{|x_{i}^{(1)}\rangle\}$ and $\{|x_{i}^{(2)}\rangle\}$
are the corresponding orthonormal basis of $\mathcal{H}_{1}$ and
$\mathcal{H}_{1}$ respectively, then
$\{|x_{i}^{(1)}\rangle\otimes|x_{j}^{(2)}\rangle\}$ is an
orthonormal basis for $\mathcal{H}_{1}\otimes\mathcal{H}_{2}$. A
general (pure) state of the composite system can be written as:

$$\rho=|\psi\rangle\langle\psi|$$

\noindent where
$|\psi\rangle=\sum_{i,j}\alpha_{ij}|x_{i}^{(1)}\rangle\otimes|x_{j}^{(2)}\rangle$.
And if $M$ represents an observable, its mean value $\langle
M\rangle$ is given by:

$$\mbox{tr}(\rho M)=\langle M\rangle$$

When observables of the form $O_{1}\otimes\mathbf{1}_{2}$ and
$\mathbf{1}_{1}\otimes O_{2}$ (with $\mathbf{1}_{1}$ and
$\mathbf{1}_{2}$ the identity operators over $\mathcal{H}_{1}$ and
$\mathcal{H}_{2}$ respectively) are considered, then partial state
operators $\rho_{1}$ and $\rho_{2}$ can be defined for systems
$S_{1}$ and $S_{2}$. The relation between $\rho$, $\rho_{1}$ and
$\rho_{2}$ is given by:
$$\rho_{1}=tr_{2}(\rho) \ \ \ \ \rho_{2}=tr_{1}(\rho)$$
\noindent where $tr_{i}$ stands for the partial trace over the $i$
degrees of freedom.  It can be shown that:
$$tr_{1}(\rho_{1}O_{1}\otimes\mathbf{1}_{2})=\langle O_{1}\rangle$$
\noindent and that a similar equation holds for $S_{2}$. Operators
of the form $O_{1}\otimes\mathbf{1}_{2}$ and $\mathbf{1}_{1}\otimes
O_{2}$ represent magnitudes related to $S_{1}$ and $S_{2}$
respectively. When $S$ is in a product state
$|\varphi_{1}\rangle\otimes|\varphi_{2}\rangle$, the mean value of
the product operator $O_{1}\otimes O_{2}$ will yield:

$$\mbox{tr}(|\varphi_{1}\rangle\otimes|\varphi_{2}\rangle\langle\varphi_{1}
|\otimes\langle\varphi_{2}|O_{1}\otimes O_{2})=\langle
O_{1}\rangle\langle O_{2}\rangle$$

\noindent reproducing statistical independence. But, as is well
known, this is not the general case.

As we pointed out above, $\rho_{1}$ and $\rho_{2}$ do not accept an
ignorance interpretation (\cite{d'esp}, \cite{Mittelstaedt},
\cite{Kirkpatrik}, \cite{ReplytoKirkpatrik}, \cite{quaternionicos},
\cite{Ufano}). Moreover, the state of the whole system
$\rho=|\psi\rangle\langle\psi|$ carries the information about the
correlations between $S_{1}$ and $S_{2}$. The fact that $\rho_{1}$
and $\rho_{2}$ are not pure states is an expression of the
non-triviality of these correlations,  that are stronger and of a
different kind than those of the classical case. This radical
difference expresses itself also in the violation of Bell
inequalities by quantum systems \cite{bell}. These facts suggest
that mixtures have to be considered as states in their own right and
be given a place in the algebraic approach to the study of quantum
properties.

\subsection{The Convex Set of States of a Quantum
System}\label{s:Convex set}

From the analysis of the last section it becomes clear that for a
complete description that includes compound systems it is not
sufficient to consider only pure states, but we have to consider
also mixtures. The standard way of doing this is by representing the
states of the system by positive, Hermitian and trace one operators,
(also called `density matrices'). The set of all density matrixes
forms a convex set (of states), which we will denote by
$\mathcal{C}$:
$$\mathcal{C}:=\{\rho\in\mathcal{A}\,|\,\mbox{tr}(\rho)=1,\,\rho\geq 0\}$$
As usual, physical observables $\mathcal{M}$ are represented by
elements $M$ of $\mathcal{A}$, the $\mathbb{R}$-vector space of
Hermitian operators acting on
$\mathcal{H}$ :%
$$\mathcal{A}:=\{ M\in B(\mathcal{H})\,|\, M=M^{\dagger} \}$$
\noindent $B(\mathcal{H})$ stands for the algebra of bounded
operators in $\mathcal{H}$. The mean value of the observable
represented by the operator $M$ when the system is in a state $\rho$
is given by $\langle M\rangle=\mbox{tr}(\rho M)$.

The set $P$ of pure states can be defined as
$$P:=\{\rho\in\mathcal{C}\,|\, \rho^{2}=\rho\}$$
This set is in correspondence with the rays of $\mathcal{H}$ by the
usual association (using Dirac notation) $[|\psi\rangle]\longmapsto
|\psi\rangle\langle\psi|$ between the elements of the projective
space of $\mathcal{H}$  and the class defined by the normalized
vector $|\psi\rangle$
($|\varphi\rangle\sim|\psi\rangle\longleftrightarrow|\varphi\rangle=\lambda|\psi\rangle$,
$\lambda\neq 0$). $\mathcal{C}$ is a convex set inside the
hyperplane $\{\rho\in\mathcal{A}\,|\,\mbox{tr}(\rho)=1\}$. If
$\dim_{\mathbb{C}}(\mathcal{H})=n<\infty$, we have an
$\mathbb{R}$-linear isomorphism $B(\mathcal{H})\cong
M_n(\mathbb{R})\times M_n(\mathbb{R})$, then
$$\mathcal{A}\cong \{(R,I)\in M_n(\mathbb{R})\times M_n(\mathbb{R})\,|\, R^t=R,\,I^t=-I\}=S_n(\mathbb{R})\times \wedge_n(\mathbb{R})$$
$$\mathcal{A}\cap\{\mbox{tr}(\rho)=1\}\cong \{(R,I)\in S_n(\mathbb{R})\times \wedge_n(\mathbb{R})\,|\,\mbox{tr}(R)=1\}$$

\noindent So the convex set $\mathcal{C}$ lies inside an
$\mathbb{R}$-algebraic variety of dimension

$$\dim_{\mathbb{R}}(\{\rho\in\mathcal{A}\,|\,\mbox{tr}(\rho)=1\})=n^2-1$$

When a system $S$ composed of subsystems $S_{1}$ and $S_{2}$ is
considered, the state of $S$ cannot be decomposed in general in a
product state $\rho=\rho_{1}\otimes\rho_{2}$, as said before. {\it
Separable states} are those states of $S$ which can be written as a
convex combination of product states \cite{GruskaImai2001}:

$$\rho_{Sep}=\sum_{k}\lambda_{k}\rho_{k}^{(1)}\otimes\rho_{k}^{(2)}$$

\noindent where $\rho_{k}^{(1)}\in\mathcal{C}_{1}$ and
$\rho_{k}^{(2)}\in\mathcal{C}_{2}$, $\sum_{k}\lambda_{k}=1$ and
$\lambda_{k}\geq 0$. It is easy to see that this expression may be
written as

$$\rho_{Sep}=\sum_{i,j}\lambda_{ij}\rho_{i}^{(1)}\otimes\rho_{j}^{(2)}$$

\noindent with $\sum_{i,j}\lambda_{ij}=1$ and $\lambda_{ij}\geq 0$.
We will denote $\mathcal{S}(\mathcal{H})$ the (convex) set of
separable states. As said above, it is a remarkable fact that there
are many states in $\mathcal{C}$ which are non separable. If the
state is non-separable, it is said to be {\it entangled}. The
estimation of the volume of $\mathcal{S}(\mathcal{H})$ is of great
interest (see for example \cite{zyczkowski1998},
\cite{horodecki2001} and \cite{aubrum2006}).

\section{The Problem of the States of the
Subsystems in QL}\label{s:explicamosporque}

In the quantum logical approach, there is a bijective correspondence
between the states of the system and the atoms of the lattice
${\mathcal{L}}_{vN}$ of its properties: the atoms of
${\mathcal{L}}_{vN}$ are the pure states. The relationship between
pure states $\rho_{pure}= |\psi\rangle\langle\psi|$ of the quantum
system and its actual properties $p$ is given by:

$$<\ |\psi\rangle\ >=\wedge\{p\in \mathcal{L}_{vN}\,|\,
p\,\,\mbox{is}\,\,\mbox{actual}\}$$

\noindent and an equivalent relation holds for the classical case.
This is an expected fact, because in ${\mathcal{L}}_{vN}$ states are
the most elemental properties of the system, up from which all other
properties are inferred. We claim that any reasonable definition of
state must satisfy this property. Furthermore, the representatives
of states must be atoms of the lattice, in order to grant that no
other non-trivial property be more elementary. But pure states form
in general a quite small subset of the border of $\mathcal{C}$ (the
atoms of $\mathcal{L}_{vN}$ are in one to one correspondence with
this subset): pure states are in a 2(N-1)-dimensional subset of the
(N$^{2}$-2)-dimensional boundary of $\mathcal{C}$. And so all
non-pure states are excluded from ${\mathcal{L}}_{vN}$. Or in the
best case, they have a different status, when considered (as in the
classical case) as measures over the lattice of projections.

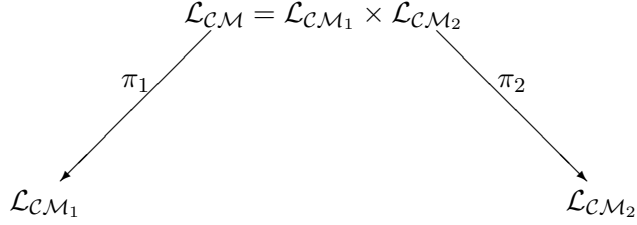
\begin{figure}\label{f:going down on classical systems}

\begin{center}
\unitlength=1mm
\begin{picture}(5,5)(0,0)
\put(-15,23){\vector(-1,-1){20}} \put(15,23){\vector(1,-1){20}}
\put(0,25){\makebox(0,0){$\mathcal{L}_{\mathcal{CM}}=\mathcal{L}_{\mathcal{CM}_{1}}\times\mathcal{L}_{\mathcal{CM}_{2}}$}}
\put(-37,0){\makebox(0,0){$\mathcal{L}_{\mathcal{CM}_{1}}$}}
\put(37,0){\makebox(0,0){$\mathcal{L}_{\mathcal{CM}_{2}}$}}
\put(-25,16){\makebox(0,0){$\pi_{1}$}}
\put(25,16){\makebox(0,0){$\pi_{2}$}}
\end{picture}
\caption{In the classical case, we can go from the state of the
system to the states of the subsystems using the set-theoretical
projections $\pi_{1}$ and $\pi_{2}$}.
\end{center}

\end{figure}

Let us emphasize taht  a remarkable problem appears in  standard QL,
linked to the status that it gives to improper mixtures (see
\cite{aerts formulation of a paradox} for more discussion on this
problem and a proposal for its solution different than the one
presented here). Suppose that $S_{1}$ and $S_{2}$ are subsystems of
a larger system $S$ which is in a pure entangled state
$|\psi\rangle$. Then we may ask which  the states of its subsystems
are. If we make the conjunction of all actual properties for, say
$S_{1}$, we will no longer obtain an atom of
$\mathcal{L}_{v\mathcal{N}_{1}}$ \cite{All properties are states}.
Instead of it, we will obtain a property which corresponds, in the
non-trivial case, to a subspace of dimension strictly greater than
one and does not correspond to the state of the subsystem. In fact,
the state of the subsystem is the (improper) mixture given by the
partial trace $\mbox{tr}_{2}(|\psi\rangle\langle\psi|)$. Thus, there
is no way to obtain the actual physical state of $S_{1}$ using the
$\wedge$ operation of $\mathcal{L}_{v\mathcal{N}_{1}}$, as it would
be reasonable according to the definition of state as minimal
property out of which all other properties are inferred.

To put things graphically, consider Figures $1$ and $2$. For the
classical case, there exist set-theoretical projections $\pi_1$ and
$\pi_2$ from $\mathcal{L}_{CM}$ to $\mathcal{L}_{CM1}$ and
$\mathcal{L}_{CM2}$ which relate the states of the system $S$ and
the states of the subsystems $S_1$ and $S_2$. In the quantum case
(Figure $2$), we do not have arrows which map states of
$\mathcal{L}_{vN}$ into states of $\mathcal{L}_{vNi}$ ($i=1,2$),
simply because non-pure states are not properly included in the
property lattice. Thus, the ``?'' arrows  of Figure $2$ are missing.

In spite of the fact that mixtures are also considered in classical
mechanics, they pose there no fundamental problem. This is so
because  classical mixtures represent a lack of information that is
-at least in principle- available. On the contrary, according to the
orthodox interpretation of $QM$, information encoded  in (improper)
mixtures is all that exists, there is no further information
available: there is no ignorance interpretation of improper
mixtures. But the orthodox quantum logical  approach puts in
different levels pure states and mixtures (the lattice of properties
and a measure over it) as is done in the classical case. In the
classical case this works, for pure states of the whole system and
its subsystems can be properly linked as Figure $1$ shows. But we
cannot do the same in the quantum case, because subsystems are
rarely found in pure states.

All of this motivates our search of algebraic structures which
contain mixtures in such a way that they may be given an equal
treatment as the one given to pure states. We will show that this is
possible and that such structures may be defined in a natural
manner, extending (in a sense explained below) $\mathcal{L}_{vN}$ so
to be compatible with the physics of compounded quantum systems.
Precisely, in the following section we construct a lattice
$\mathcal{L}$ that has all physical states as its atoms and whose
meet operation over all actual properties of a system gives the
actual physical state of that system. It also includes
$\mathcal{L}_{vN}$ set theoretically, so we are able to reobtain all
well known results of single isolated systems.

\begin{figure}\label{f:tracesdonotwork}
\begin{center}
\unitlength=1mm
\begin{picture}(5,5)(0,0)
\put(-3,23){\vector(-1,-1){20}} \put(3,23){\vector(1,-1){20}}
\put(0,25){\makebox(0,0){$\mathcal{L}_{v\mathcal{N}}$}}
\put(-27,0){\makebox(0,0){${\mathcal{L}_{v\mathcal{N}_{1}}}$}}
\put(27,0){\makebox(0,0){${\mathcal{L}_{v\mathcal{N}_{2}}}$}}
\put(-15,16){\makebox(0,0){$?$}} \put(15,16){\makebox(0,0){$?$}}
\end{picture}
\caption{We cannot apply partial traces in order to go down from
$\mathcal{L}_{v\mathcal{N}}$ to $\mathcal{L}_{v\mathcal{N}_{1}}$,and
$\mathcal{L}_{v\mathcal{N}_{2}}$.}
\end{center}
\end{figure}
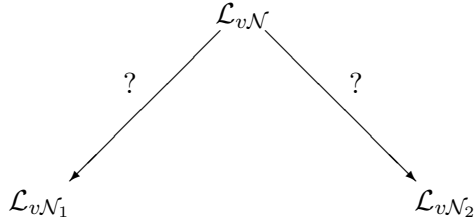

$\mathcal{L}$ is constructed in such a way that there exist
projection functions which map \emph{all} states (atoms) of the
structure corresponding to the whole system $S$ to the corresponding
states (atoms) of its subsystems $S_{1}$ and $S_{2}$. This
assignation rule is compatible with the physics of the problem,
i.e., it is constructed using partial traces, which are the natural
functions which map states of the larger system with the states of
it subsystems. Improper mixtures are put in correspondence with
atoms of $\mathcal{L}$, granting that they are the most elementary
properties.

There is another important feature of $\mathcal{L}$. In
$\mathcal{L}_{v\mathcal{N}}$  from two given pure states, say
$|\psi_{1}\rangle$ and $|\psi_{2}\rangle$, a new state
$\alpha|\psi_{1}\rangle+\beta|\psi_{2}\rangle$ may be constructed;
we have at hand the superposition principle. The
$\vee_{\mathcal{L}_{v\mathcal{N}}}$ operation of the von Neumann
lattice is directly linked to the superposition principle: starting
with two rays, the $\vee_{\mathcal{L}_{v\mathcal{N}}}$ operation
yields the closed subspace formed by all linear combinations of the
generators of the rays. But there is another operation available,
namely we can mix states, we can perform a ``mixing operation'' to
get
$p_{1}|\psi_{1}\rangle\langle\psi_{1}|+p_{2}|\psi_{2}\rangle\langle\psi_{2}|$.
There is no place for such a thing in $\mathcal{L}_{v\mathcal{N}}$,
but it may be performed in $\mathcal{L}$. The $\vee_{\mathcal{L}}$
operation reflects the fact that we can mix states, playing an
analogous role to that of $\vee_{\mathcal{L}_{v\mathcal{N}}}$ in
relation to the superposition principle.

Let us mention before presenting $\mathcal{L}$ that there exists a
trivial example of a lattice that fulfills the requirement that its
atoms are improper mixtures, namely, the set of all subsets of
$\mathcal{C}$, which we call $\mathcal{P}(\mathcal{C})$. If we use
set intersection as the meet operation and set union as the join
operation, this structure is a Boolean lattice. But this lattice is
not of interest, because its disjunction is not connected with the
mixing operation mentioned above, alike $\vee_{\mathcal{L}}$. Its
boolean structure hides the radical differences between classical
and quantum mechanics. But this trivial example shows that our
construction below may be one among a family of possible lattices
which overcome the problem of $\mathcal{L}_{v\mathcal{N}}$ presented
in this section.

\section{The Lattice of Density Operators}\label{s:New language}

In order to construct a lattice for density operators, let us
consider the pair $G(\mathcal{A}):=(\mathcal{A},\mbox{tr})$ where
$\mathcal{A}$ is the $\mathbb{R}$-vector space of operators over
$\mathcal{H}$ and $\mbox{tr}$ is the usual trace operator on
$B(\mathcal{H})$, which induces the scalar product $<A,B>=tr(A\cdot
B)$ ($\dim(\mathcal{H})<\infty$). The restriction to $\mathcal{A}$
of $\mbox{tr}$, makes $\mathcal{A}$ into an $\mathbb{R}$-Euclidean
vector space. With the standard $\vee$, $\wedge$ and $\neg$
operations, $G(\mathcal{A})$ is a modular, orthocomplemented, atomic
and complete lattice (not distributive, hence not a Boolean
algebra).

Let $\mathcal{L}_o$ be the set of subspaces:
$$\mathcal{L}_o:=\{L = S\cap\mathcal{C}\,|\, S\in G(\mathcal{A})\}$$
\noindent There are a lot of subspaces $S, S_i \in G(\mathcal{A})$
such that $S\cap {\mathcal{C}}=S_i\cap \mathcal{C}$, so for each
$L\in\mathcal{L}_o$ we may choose the subspace with the least
dimension $[S]$ as the representative element:
$$[S]:=\min\{\dim_{ \mathbb{R} }(S)\,|\,L=S\cap\mathcal{C},\,S\in G(\mathcal{A})\}$$
\noindent Let $[S]=L$, being  $S\in G(\mathcal{A})$ an element of
the class $L$, then
$$S\cap\mathcal{C}\subseteq<S\cap\mathcal{C}>_{\mathbb{R}}\subseteq
S\Rightarrow
S\cap\mathcal{C}\cap\mathcal{C}\subseteq<S\cap\mathcal{C}>_{\mathbb{R}}\cap\mathcal{C}
\subseteq S\cap\mathcal{C}\Rightarrow$$
$$<S\cap\mathcal{C}>\cap\mathcal{C}=S\cap\mathcal{C}$$
\noindent So $<S\cap\mathcal{C}>$ and $S$ are in the same class $L$.
Note that $<S\cap\mathcal{C}>\subseteq S$ and if $S$ is the subspace
with the least dimension, then $<S\cap\mathcal{C}>=S$. Also note
that the representative with least dimension is unique, because if
we choose $S'$ such that $S'\cap\mathcal{C}=S\cap\mathcal{C}$, then
$$S=<S\cap\mathcal{C}>=<S'\cap\mathcal{C}>=S'$$
\noindent Finally, the representative of a class $L$ that we choose
is the unique $\mathbb{R}$-subspace $S\subseteq\mathcal{A}$ such
that
$$S=<S\cap\mathcal{C}>_{\mathbb{R}}$$
\noindent We call it the {\it good representative}. It is important
to remark that in the case of infinite dimensional Hilbert spaces we
cannot define good representatives in such a way.

Let us now define $\vee$, $\wedge$ and $\neg$ operations in
$\mathcal{L}_o$ as:
$$(S\cap\mathcal{C}) \wedge (T\cap\mathcal{C})=
<S\cap\mathcal{C}>\cap<T\cap\mathcal{C}>\cap\mathcal{C}$$
$$(S\cap\mathcal{C}) \vee (T\cap\mathcal{C})=
(<S\cap\mathcal{C}>+<T\cap\mathcal{C}>)\cap\mathcal{C}$$
$$\neg(S\cap\mathcal{C})=<S\cap\mathcal{C}>^\perp\cap\mathcal{C}$$
\noindent They are well defined for every element of the classes
$[S]$ and $[T]$. It is easy to see that $\mathcal{L}=<\mathcal{L}_o,
\vee, \wedge,\mathbf{0},\ \mathbf{1}> $ is a complete lattice, with
$\mathbf{0}=\emptyset$ represented by the class of $G(\mathcal{A})$
whose elements are disjoint with $\mathcal{C}$ and
$\mathbf{1}=\mathcal{C}$, represented by the class of $\mathcal{A}$.
It is an atomic lattice: the atoms of $\mathcal{L}$ are given by the
intersection of rays in $G(\mathcal{A})$ and $\mathcal{C}$. They are
the sets $\{\rho\}$, with $\rho$ a density operator.

It is important to notice that with respect to the $\neg$ operation,
$\mathcal{L}$ is not an orthocomplemented lattice -alike
$\mathcal{L}_{v\mathcal{N}}$- because if we take
$L=\{\frac{1}{N}\mathrm{1}\}=<\frac{1}{N}\mathrm{1}>\cap\mathcal{C}$,
then

$$\neg(\neg L)=\neg(<\frac{1}{N}\mathrm{1}>^{\perp}\cap\mathcal{C})=\neg\emptyset=\mathcal{C}\neq L$$

\noindent On the other hand it is easy to show that
non-contradiction holds

$$L\wedge\neg L=\mathbf{0}$$

\noindent and also contraposition

$$L_{1}\leq L_{2}\Longrightarrow \neg L_{2}\leq \neg L_{1}$$

\begin{prop}
If $\dim(\mathcal{H})<\infty$, $\mathcal{L}$ is a modular lattice.
\end{prop}
\begin{proof}
\noindent To prove the modular equality
$$[S]\leq [R] \Rightarrow [S]\vee ([T]\wedge [R])=([S]\vee [T])\wedge [R]$$
the key point is that
$$[S]\leq [R]\Leftrightarrow S\cap\mathcal{C}\subseteq R\cap\mathcal{C}\Rightarrow S=<S\cap\mathcal{C}>\subseteq<R\cap\mathcal{C}>=R$$
\noindent So, using $S\subseteq R$, is easy to see that $(S+(T\cap
R))\cap\mathcal{C}=((S+T)\cap R)\cap\mathcal{C}$.
\end{proof}

Furthermore, we can prove the following:
\begin{prop}
There is a one to one correspondence between the states of the
system and the atoms of $\mathcal{L}$.
\end{prop}
\begin{proof}
For every $\rho\in$ $\mathcal{C}$, we have that $<\rho>\cap \
\mathcal{C}=\{\rho\}$. This is so because the only positive matrix
of trace one that is a multiple of $\rho$ is $\rho$ itself. Then,
$\{\rho\}$ is an element of $\mathcal{L}$. Suppose that there exists
$L$ such that $\mathbf{0}\leq L\leq \{\rho\}$. If $L\neq\mathbf{0}$,
we can write $L=S\cap\mathcal{C}$, with $S$ being the good
representative for the class of $L$. $L\leq \{\rho\}$ implies that
$S\subseteq<\rho>$ and thus $S=<\rho>$, so it follows that
$L=\{\rho\}$. Conversely, if $L$ is an atom of $\mathcal{L}$, take
$\rho\in L$. Define $L'=<\rho>\cap\ \mathcal{C}=\{\rho\}$. It is
clear that $L'\subseteq L$ and, given that $L'\neq\mathbf{0}$, we
have $L'=L$.
\end{proof}

The last proposition shows that we can represent the states of
subsystems of a larger system as elements of the lattice
$\mathcal{L}$ giving them a similar status as pure states, something
impossible in the standard formalism of QL and one of the desiderata
in searching a structure to deal with composite systems. It is a
well established fact \cite{bengtssonyczkowski2006}, that there is a
lattice isomorphism between the complemented and complete lattice of
faces of the convex set $\mathcal{C}$ and $\mathcal{L}_{vN}$. As
desired, ${\mathcal{L}}_{vN}$ is included in $\mathcal{L}$
guaranteing to represent  all the good features of standard QL in
the new algebra. This is a non trivial result, and it is ensured by
the following proposition and its corollary:
\begin{prop}
Every face of $\mathcal{C}$ is an element of $\mathcal{L}$.
\end{prop}
\begin{proof}
Let $F\subseteq \mathcal{C}$ be a face. Then there exists a
$\mathbb{R}$-hyperplane $H$
inside $\{\rho\in\mathcal{A}\,|\,\mbox{tr}(\rho)=1\}$ such that $F=H\cap\mathcal{C}$.\\
Given that $H=\{ l=\alpha\}$ with $\alpha\in\mathbb{R}$ and $l$ an
$\mathbb{R}$-linear form on $\mathcal{A}$, we have that:
$$F=H\cap\mathcal{C}=H\cap\mathcal{C}\cap\{\mbox{tr}=1\}=
\{l=\alpha,\,\mbox{tr}=1\}\cap\mathcal{C}=$$
$$\{l=\alpha\mbox{tr},\,\mbox{tr}=1\}\cap\mathcal{C}=
\{l=\alpha\mbox{tr}\}\cap\mathcal{C}\cap\{\mbox{tr}=1\}=
\{l-\alpha\mbox{tr}=0\}\cap\mathcal{C}$$ So
$\{\rho\in\mathcal{A}\,|\,l(\rho)-\alpha\mbox{tr}(\rho)=0\}\in
G(\mathcal{A})$, and then $F\in \mathcal{L}$.
\end{proof}

So, we can naturally embed $\mathcal{L}_{v\mathcal{N}}$ in
$\mathcal{L}$ as a poset.

\begin{coro}
The complemented and complete lattice of faces of the convex set
$\mathcal{C}$ is a subposet of $\mathcal{L}$.
\end{coro}
\begin{proof}
We have already seen that $\mathcal{L}_{vN}\subseteq\mathcal{L}$ as
sets. Moreover it is easy to see that if $F_1\leq F_2$ in
$\mathcal{L}_{vN}$ then $F_1\leq F_2$ in $\mathcal{L}$. This is so
because both orders are set theory inclusions.
\end{proof}

The previous Corollary shows that $\mathcal{L}$ and
$\mathcal{L}_{vN}$ are closely connected. Let us analyze the
relationship between the operations of the two lattices in order to
characterize this connection. We recall that the meet of two faces
is their intersection and the join is the smallest face containing
both. In $\mathcal{L}_{vN}$, the meet of two subspaces is their
intersection and the join is their closed linear spam.
\begin{enumerate}

\item[$\wedge$:]
$F_1,F_2\in\mathcal{L}_{vN}$, then $F_1\wedge F_2$ in
$\mathcal{L}_{vN}$ is the same as in $\mathcal{L}$. So the inclusion
$\mathcal{L}_{vN}\subseteq\mathcal{L}$ preserves the
$\wedge$-operation.

\item[$\vee$:]
In general it does not preserve the $\vee$-operation. The relation
between the two operations is:
$$F_1\vee_{\mathcal{L}} F_2\leq F_1\vee_{\mathcal{L}_{vN}} F_2$$
$$F_1\leq F_2\Rightarrow F_1\vee_{\mathcal{L}} F_2=F_1\vee_{\mathcal{L}_{vN}} F_2=F_2$$
For example, if the convex set $\mathcal{C}$ is a rectangle and
$F_1$ and $F_2$ are two opposite vertices then, the face-join of
them is the whole rectangle, and the $\mathcal{L}$-join is the
diagonal joining them.

\item[$\neg$:]
In any lattice, $x$ is a complement to $y$ if $x\vee y=1$ and
$x\wedge y=0$. In general the lattice of faces of a convex set is
complemented, but in the case of $\mathcal{C}$ it is
orthocomplemented, that is, it has a distinguished complemented face
for every face $F\subset\mathcal{C}$. Given that
$\mathcal{L}_{vN}\cong\mathcal{P}(\mathcal{H})$, the lattice of
projectors in $\mathcal{H}$, the $\neg$-operation in
$\mathcal{L}_{vN}$ is that induced from $\mathcal{P}(\mathcal{H})$.
If $F\subseteq\mathcal{C}$ is a face, there exists a unique
projector $P\in \mathcal{A}$ such that
$$F=\{\rho\in\mathcal{C}\,|\, \mbox{tr}(P\rho)=0\}=\{\rho\in\mathcal{C}\,|\,\rho\perp P\}
\Rightarrow$$
$$\neg_{\mathcal{L}_{vN}} F=\{\rho\in\mathcal{C}\,|\, \rho\perp 1-P\}$$
It is easy (using eigenvalues) to see that it is well defined and
that $\neg F$ is again a face. Given that $F\in\mathcal{L}$, it has
a good representative $F=[S]$. Then
$$\neg_{\mathcal{L}} F=S^\perp\cap\mathcal{C}$$
Using this, we can prove that $\neg_{\mathcal{L}}F\leq
\neg_{\mathcal{L}_{vN}}F$, because:
$$\nu\in\neg_{\mathcal{L}}F \mbox{ then }\nu\perp\rho, \, \rho\in F$$
and, in particular,
$$\nu\perp 1-P\mbox{ then }\nu\in \neg_{\mathcal{L}_{vN}}F$$
\end{enumerate}

\subsection{Quantum Interactions Enlarge the Lattice of
Properties}\label{s:The Relationship}

The results of the last section show that $\mathcal{L}$ is a quite
natural extension of $\mathcal{L}_{vN}$ and satisfies that improper
mixtures are in a bijective correspondence with the atoms of the
lattice. This feature allows this lattice to avoid the problems
which appear in the standard formulation of QL posed in \cite{aerts
formulation of a paradox} and also discussed in section
\ref{s:explicamosporque} of this work. In the new lattice, the
conjunction of all actual properties  yields the physical state of
the system because all states are in correspondence with atoms,
which are minimal elements. From the physical point of view the
necessity of an extension becomes clear from the comparison between
classical and quantum compound systems. When we have a single
classical system, its properties are faithfully represented by the
subsets of its phase space. When another classical system is added
and the compound system is considered, no enrichment of the state
space of the former system is needed in order to describe its
properties, \emph{even} in the presence of interactions. No matter
which the interaction may be, the cartesian product of the
individual phase spaces gives all is needed to represent the
compound system, and the same stands for the property lattices. But
the situation is quite different in quantum mechanics. This is so
because if we add a new quantum system to the first one, pure states
are no longer faithful in order to describe subsystems. Interactions
produce non trivial correlations, which are reflected in the
presence of entangled states and violation of Bell inequalities.
These non trivial correlations are behind the fact that within the
standard quantum logical approach, the conjunction of all actual
properties does not yield the physical state of the subsystem. Thus,
besides their own properties, we need information about the non
trivial \emph{correlations} that each subsystem has with other
subsystems -for example, a system with the environment- that may be
regarded as new elements in the structure of properties and cannot
be described otherwise. For this reason an enlargement of the
lattice of properties is needed to represent improper mixtures by
atoms $\{\rho\}$ in $\mathcal{L}$. We will come back to this point
in Subsection \ref{s:projections}, where we study the projections
from the lattice of the compound system onto the lattices of the
subsystems.

\section{The Relationship Between $\mathcal{L}$ and ${\mathcal{L}}_i$}\label{s:The Relationship}

Given two systems with Hilbert spaces $\mathcal{H}_{1}$ and
$\mathcal{H}_{2}$, we can construct the lattices $\mathcal{L}_{1}$
and $\mathcal{L}_{2}$ according to the procedure of section
\ref{s:New language}. We can also construct $\mathcal{L}$, the
lattice associated to the product space
$\mathcal{H}_{1}\otimes\mathcal{H}_{2}$. In this section we examine
their mutual relations. We study some special maps between these
lattices and their properties, in order to get an insight in the
characterization of compound quantum systems.

\subsection{Separable States (Going Up)}\label{s:going up}

We start defining the map:

$$\Psi:\mathcal{L}_{1}\times\mathcal{L}_{2}\longrightarrow\mathcal{L}$$
$$(S_{1}\cap\mathcal{C}_{1},S_{2}\cap\mathcal{C}_{2})\longrightarrow S\cap\mathcal{C}$$
$$\mbox{where }S=(<S_{1}\cap\mathcal{C}_{1}>\otimes<S_{2}\cap\mathcal{C}_{2}>)$$

\noindent In terms of good representatives,
$\Psi([S_1],[S_2])=[S_1\otimes S_2]$. We can prove the following:

\begin{prop}\label{bimorfism}
Fixing $[U]\in\mathcal{L}_2$ then $\mathcal{L}_{1}$ is isomorphic
(as complete lattice) to $\mathcal{L}_{1}\times[U] \subseteq
\mathcal{L}$. The same is true for $\mathcal{L}_{2}$ and an
arbitrary element of $\mathcal{L}_1$.
\end{prop}\label{bi-mor}
\begin{proof}
Let us prove it for $\mathcal{L}_1$.  Let
$([S],[U])\in\mathcal{L}_{1}\times[U]$ with $S$ a good
representative for $[S]$ and $U$ for $[U]$. When we apply $\Psi$ we
obtain the proposition $[S\otimes U]\in\mathcal{L}$, then, we can
consider the image under $\Psi$ of $\mathcal{L}_{1}\times
[U]\subseteq\mathcal{L}_{1}\times\mathcal{L}_{2}$:
$$\Psi(\mathcal{L}_{1}\times[U])=\{[S\otimes U]\mbox{ where }S
\mbox{ is a good representative for } [S]\in\mathcal{L}_{1}\}$$

\noindent From this characterization it is easy to see that $\Psi$
is injective. If $[S\otimes U]=[T\otimes U]$ ($S$ and $T$ are good
representatives), taking partial traces (more in Subsection
\ref{s:projections}) then $[S]=[T]$.

Moreover, $\Psi(-,[U])$ is a lattice morphism: let $[S\otimes
U],\,[T\otimes U]\in\mathcal{L}$ with $S$ and $T$ good
representatives of $[S],[T]\in\mathcal{L}_1$. The key observation is
that $S\otimes U$ and $T\otimes U$ are also good representatives
(taking partial traces). Then we have:

$$[S\otimes U]\wedge[T\otimes U]=[(S \cap T)\otimes U]=\Psi([S]\wedge[T],[U])$$
$$[S\otimes U]\vee[T\otimes U]=[(S \oplus T)\otimes U]=\Psi([S]\vee[T],[U])$$

\noindent This ensures that $\mathcal{L}_{1}$ is a sublattice of
$\mathcal{L}$. The same is true for $\mathcal{L}_{2}$.
\end{proof}

Notice that we can use an arbitrary atom $\rho_2\in\mathcal{C}_2$
instead of some $[U]\in\mathcal{L}_2$ and that the application
$\Psi$ restricted to $\mathcal{L}_{1}\times \rho_2$ does not
preserve the $\neg$-operation. This is so, because:

$$\Psi(\neg[S],\rho_2)=[S^\perp
\otimes \rho_2]\subset[(S\otimes \rho_2)^\perp]=$$
$$=\neg[S\otimes \rho_2]=\neg\Psi([S],\rho_2)$$

\noindent The inclusion holds, because if $\rho\in [S^\perp
\otimes\rho_2]$, then $\rho=(\Sigma\lambda_{i}\rho_{i})\otimes
\rho_2=\Sigma\lambda_{i}\rho_{i}\otimes \rho_2$, with $\rho_{i}\in
S^\perp$. It is clear that all the $\rho_{i}\otimes\rho_2$ are
orthogonal to $S\otimes\rho_2$, and then $\rho\in
(S\otimes\rho_2)^\perp$. In general the inclusion is strict, because
we can have elements of the form $\rho_{1}\otimes\rho'_{2}$, with
$\rho_{1}\in S^\perp$ and $\rho'_{2}\neq\rho_2$. Then,
$\rho_{1}\otimes\rho_{2}\in S\otimes\rho_2$, but
$\rho_{1}\otimes\rho_{2}\notin S^\perp \otimes \rho_2$. This has a
clear physical meaning: in fact, when the system $S_{1}$ is
isolated, its lattice of properties $\mathcal{L}_{1}$ is equivalent
to $\mathcal{L}_{1}\times\rho_2$. But when we add system $S_{2}$ we
can, for example, prepare the systems independently, in such a way
that the state after preparation is $\rho_{1}\otimes\rho'_{2}$ with
$\rho_{1}\in S^\perp$ and $\rho'_{2}$ an arbitrary state of $S_{2}$.
Then, we see that there is much more freedom in the space of all
states.

Let us study now the image of $\Psi$. First, we note that given
$L_{1}\in\mathcal{L}_{1}$ and $L_{2}\in\mathcal{L}_{2}$, we can
define the following convex tensor product:
$$L_{1}\widetilde{\otimes}\,L_{2}:=\{\sum\lambda_{ij}\rho_{i}^{1}\otimes\rho_{j}^{2}\,|\,\rho_{i}^{1}
\in L_{1},\,\,\rho_{j}^{2}\in L_{2},\,\, \sum\lambda_{ij}=1
\,\,\mbox{and} \,\,\lambda_{ij}\geq 0\}$$ This product is formed by
all possible convex combinations of tensor products of elements of
$L_{1}$ and elements of $L_{2}$, and it is again a convex set.

\begin{prop}
$L_{1}\widetilde{\otimes}\,L_{2}\subseteq\Psi(L_1,L_2)$
\end{prop}

\begin{proof}
If $\rho\in L_{1}\widetilde{\otimes}\,L_{2}$, then
$\rho=\sum\lambda_{ij}\rho_{i}^{1}\otimes\rho_{j}^{2}$, with
$\rho_{i}^{1}\in L_{1}$, $\rho_{j}^{2}\in L_{2}$,
$\sum\lambda_{ij}=1$ and $\lambda_{ij}\geq 0$. For each $i,j$,
$\rho_{i}^{1}\otimes\rho_{j}^{2}$ is again a positive trace one
operator and so belongs to $\mathcal{C}$. It belongs to
$<L_{1}>\otimes<L_{2}>$ because of the definition of tensor product.
Then, it belongs to $\Psi(L_1,L_2)$. As $\mathcal{C}$ is convex,
then $\rho\in\mathcal{C}$, because it is a convex combination of
elements in $\mathcal{C}$. It is a linear combination of elements of
$<L_{1}>\otimes<L_{2}>$ also, and so it belongs to it. This proves
that $\rho\in\Psi(L_1,L_2)$.
\end{proof}

\noindent We can also prove that:

\begin{prop}
If $L\in Im(\Psi)$, then $L\cap
\mathcal{S}(\mathcal{H})\neq\emptyset$.
\end{prop}

\begin{proof}
$L\in Im(\Psi)$ implies that there exist $L_{1}$ and $L_{2}$ such
that $L=\Psi(L_{1},L_{2})$. By definition
$\Psi(L_{1},L_{2})=(S_{1}\otimes S_{2})\cap\mathcal{C}$, with
$L_{1}=S_{1}\cap\mathcal{C}_{1}$ and
$L_{2}=S_{2}\cap\mathcal{C}_{2}$. Let $\rho_{1}\in L_{1}$ and
$\rho_{2}\in L_{2}$. Then, $\rho_{1}\otimes\rho_{2}\in L$. But we
have also that $\rho_{1}\otimes\rho_{2}\in
\mathcal{S}(\mathcal{H})$, and then $L\cap
\mathcal{S}(\mathcal{H})\neq\emptyset$.
\end{proof}

From the last proposition it follows that $Im(\Psi)\subset
\mathcal{L}$, because if we take a nonseparable state
$\rho\in\mathcal{C}$, then ${\rho}\in \mathcal{L}$, but
${\rho}\cap\mathcal{S}(\mathcal{H})=\emptyset$, and so, it cannot
belong to $Im(\psi)$. Note that in general
$L_{1}\widetilde{\otimes}\,L_{2}$ is not an element of
$\mathcal{L}$.

Let us compute
$\mathcal{C}_{1}\widetilde{\otimes}\,\mathcal{C}_{2}$. Remember that
$\mathcal{C}_{1}=[\mathcal{A}_{1}]\in\mathcal{L}_{1}$ and
$\mathcal{C}_{2}=[\mathcal{A}_{2}]\in\mathcal{L}_{2}$:
$$\mathcal{C}_{1}\widetilde{\otimes}\,\mathcal{C}_{2}=
\{\sum\lambda_{ij}\rho_{i}^{1}\otimes\rho_{j}^{2}\,|\,\rho_{i}^{1}
\in \mathcal{C}_{1},\,\,\rho_{j}^{2}\in \mathcal{C}_{2},\,\,
\sum\lambda_{ij}=1 \,\,\mbox{and} \,\,\lambda_{ij}\geq 0\}$$

\noindent So, using the definition of $\mathcal{S(\mathcal{H})}$,
the set of all separable states, we have

$$\mathcal{S(\mathcal{H})}=\mathcal{C}_{1}\widetilde{\otimes}\,\mathcal{C}_{2}$$

\noindent We know that
$\mathcal{C}_{1}\widetilde{\otimes}\,\mathcal{C}_{2}\subset\mathcal{C}$.
But it does not necessarily belong to $\mathcal{L}$. We can prove
also the following propositions:

\begin{prop}
Let $L\in Im(\Psi)$ and $\rho\in L$. Then, $\rho$ is a linear
combination of product states.
\end{prop}

\begin{proof}
Let $L\in Im(\Psi)$. Then, there exist $L_{1}\in\mathcal{L}_{1}$ and
$L_{2}\in\mathcal{L}_{2}$ such that $\Psi(L_{1},L_{2})=L$. If
$L_{1}=S_{1}\cap\mathcal{C}_{1}$ and
$L_{2}=S_{2}\cap\mathcal{C}_{2}$, with $S_{1}$ and $S_{2}$ good
representatives, we have:

$$L=(S_{1}\otimes S_{2})\cap\mathcal{C}\Longrightarrow
\rho=\sum_{i,j}\lambda_{ij}\rho_{i}^{1}\otimes\rho_{j}^{2}$$
\end{proof}

\begin{prop}
Let $\rho=\rho_{1}\otimes\rho_{2}$, with
$\rho_{1}\in\mathcal{C}_{1}$ and $\rho_{2}\in\mathcal{C}_{2}$. Then
$\{\rho\}=\Psi(\{\rho_{1}\},\{\rho_{2}\})$ with
$\{\rho_1\}\in\mathcal{L}_1$, $\{\rho_2\}\in\mathcal{L}_2$ and
$\{\rho\}\in\mathcal{L}$.
\end{prop}

\begin{proof}
We already know that the atoms are elements of the lattices.
$$\Psi(\{\rho_{1}\},\{\rho_{2}\})
=(<\rho_{1}>\otimes<\rho_{2}>)\cap\,\,\mathcal{C}=
<\rho_{1}\otimes\rho_{2}>\cap\,\,\mathcal{C}=\{\rho_{1}\otimes\rho_{2}\}=\{\rho\}$$
\end{proof}

\begin{prop}
Let $\rho\in\mathcal{S(\mathcal{H})}$, the set of separable states.
Then, there exist $L\in\mathcal{L}$, $L_{1}\in\mathcal{L}_{1}$ and
$L_{2}\in\mathcal{L}_{2}$ such that $\rho\in L$ and
$L=\Psi(L_{1},L_{2})$.
\end{prop}

\begin{proof}
If $\rho\in\mathcal{S(\mathcal{H})}$, then
$\rho=\sum_{ij}\lambda_{ij}\rho_{i}^{1}\otimes\rho_{j}^{2}$, with
$\sum_{ij}\lambda_{ij}=1$ and $\lambda_{ij}\geq 0$. Consider the
subspaces:

$$S_{1}=<\rho_{1}^{1},\rho_{2}^{1},\cdots,\rho_{k}^{1}>\quad
S_{2}=<\rho_{1}^{2},\rho_{2}^{2},\cdots,\rho_{l}^{2}>$$

\noindent Take $L_{1}=S_{1}\cap\mathcal{C}_{1}$ and
$L_{2}=S_{2}\cap\mathcal{C}_{2}$. Let us observe first that
$<S_{1}\cap\mathcal{C}_{1}>\subseteq S_{1}$. We have
$\rho_{i}^{1}\in\mathcal{C}_{1}$ and so,
$<S_{1}\cap\mathcal{C}_{1}>= S_{1}$, because $S_{1}$ is generated by
the set {$\rho_{i}^{1}$}. We also have that
$<S_{2}\cap\mathcal{C}_{2}>= S_{2}$. Now we can compute:
$$\Psi(L_{1},L_{2})=(<S_{1}\cap\mathcal{C}_{1}>\otimes<S_{2}\cap\mathcal{C}_{2}>)\cap\mathcal{C}=
(S_{1}\otimes S_{2})\cap\mathcal{C}$$ \noindent But the set
$\{\rho_{i}^{1}\otimes\rho_{j}^{2}\}$ generates $S_{1}\otimes
S_{2}$, and then, $(S_{1}\otimes S_{2})\cap\mathcal{C}$ is formed by
all the possible convex combinations of
$\{\rho_{i}^{1}\otimes\rho_{j}^{2}\}$. This proves that $\rho\in L$.
\end{proof}

The above propositions show that $Im(\Psi)$ encodes information
related to separable states. As a general state in $S$ is non
separable, we obtain that $Im(\Psi)$ is not equal to $\mathcal{L}$.
This is a reasonable result. If we interpret $\mathcal{L}_{1}$ and
$\mathcal{L}_{2}$ as encoding all the information that is available
for $S_{1}$ and $S_{2}$ expressed via observables of the subsystems
separately,  it will never be possible to reconstruct from it alone
all the information about the correlations between $S_{1}$ and
$S_{2}$, which is encoded in $\mathcal{L}$. This information is
available only in observables of the whole system $S$. From
$Im(\Psi)$ it is possible to recover information about separated
states only. As said above, the tensor product contains  more
information than that of its parties, and this is directly linked to
the non existence of a satisfactory theory of tensor products of
orthomodular posets and lattices compatible with physics.

\subsection{Projections Onto $\mathcal{L}_{1}$ and $\mathcal{L}_{2}$ (Going
Down)}\label{s:projections}

There are other maps of interest. If the whole system is in a state
$\rho$, using partial traces we can define states for the subsystem
$\rho_{1}=tr_{2}(\rho)$ and similarly for $\rho_{2}$. Then, we can
consider the maps:

$$\mbox{tr}_{i}:\mathcal{C}\longrightarrow \mathcal{C}_{j}
\quad|\quad \rho\longrightarrow \mbox{tr}_{i}(\rho)$$

\noindent from which we can construct the induced projections:

$$\tau_{i}:\mathcal{L}\longrightarrow \mathcal{L}_{i}
\quad|\quad S\cap\mathcal{C}\longrightarrow \mbox{tr}_{i}(
<S\cap\mathcal{C}> )\cap \mathcal{C}_i$$ In terms of good
representatives $\tau_i([S])=[\mbox{tr}_i(S)]$. Then we can define
the product map

$$\tau:\mathcal{L}\longrightarrow\mathcal{L}_{1}\times\mathcal{L}_{2}
\quad|\quad L\longrightarrow(\tau_{1}(L),\tau_{2}(L))$$ \noindent We
can prove the following about the image of $\tau_{i}$.

\begin{prop}\label{lastausonsurjective}
The functions $\tau_{i}$ are surjective and preserve the
$\vee$-operation. They are not injective.
\end{prop}

\begin{proof}
Take $L_{1}\in\mathcal{L}_{1}$. Choose an arbitrary element of
$\mathcal{C}_{2}$, say $\rho_{2}$. Now consider the following
element of $\mathcal{L}$

$$L=<L_{1}\otimes\rho_{2}>\cap\mathcal{C}$$

\noindent It is clear that $\tau_{1}(L)=L_{1}$, because if
$\rho_{1}\in L_{1}$, then
$\mbox{tr}(\rho_{1}\otimes\rho_{2})=\rho_{1}$. So, $\tau_{1}$ is
surjective. On the other hand, the arbitrariness of $\rho_{2}$
implies that it is not injective. An analogous argument follows for $\tau_2$.\\
Let us see that $\tau_i$ preserves the $\vee$-operation:

$$\tau_i([S]\vee[T])=\tau_i([S\oplus T])=[\mbox{tr}_i(S\oplus T)]=$$
$$=[\mbox{tr}_i(S)\oplus \mbox{tr}_i(T)]=
[\mbox{tr}_i(S)]\vee [\mbox{tr}_i(T)]=\tau_i([S])\vee \tau_i([T])$$
\end{proof}

\noindent Let us now consider the $\wedge$-operation. Let us
compute:

$$\tau_i([S]\wedge[T])=\tau_i([S\cap T])=[\mbox{tr}_i(S\cap T)]\subseteq$$
$$\subseteq[\mbox{tr}_i(S)\cap \mbox{tr}_i(T)]=
[\mbox{tr}_i(S)]\wedge [\mbox{tr}_i(T)]=\tau_i([S])\wedge
\tau_i([T])$$

\noindent It is easy to see that $\mbox{tr}_i(S\cap
T)\subseteq\mbox{tr}_i(S)\cap \mbox{tr}_i(T)$. This is because if
$\rho\in\mbox{tr}_i(S\cap T)$, then $\rho=\mbox{tr}_i(\sigma)$, with
$\sigma\in S$ and  $\sigma\in T$. This means that
$\rho\in\mbox{tr}_i(S)\cap \mbox{tr}_i(T)$, and so we have the
inclusion of classes. But these sets are not equal in general, as
the following example shows. Take
$\{\rho_{1}\otimes\rho_{2}\}\in\mathcal{L}$ and
$\{\rho_{1}\otimes\rho'_{2}\}\in\mathcal{L}$, with $\rho'\neq\rho$.
It is clear that
$\{\rho_{1}\otimes\rho_{2}\}\wedge\{\rho_{1}\otimes\rho'_{2}\}=\mathbf{0}$
and so,
$\tau_{1}(\{\rho_{1}\otimes\rho_{2}\}\wedge\{\rho_{1}\otimes\rho'_{2}\})=\mathbf{0}$.
On the other hand,
$\tau_{1}(\{\rho_{1}\otimes\rho_{2}\})=\{\rho_{1}\}=\tau_{1}(\{\rho_{1}\otimes\rho'_{2}\})$,
and  so,
$\tau_{1}(\{\rho_{1}\otimes\rho_{2}\})\wedge\tau_{1}(\{\rho_{1}\otimes\rho'_{2}\})=\{\rho_{1}\}$.
A similar fact holds for the $\neg$-operation.

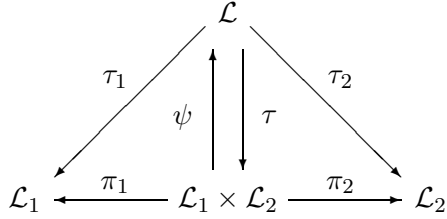
\begin{figure}\label{f:maps}
\begin{center}
\unitlength=1mm
\begin{picture}(5,5)(0,0)
\put(-3,23){\vector(-1,-1){20}} \put(3,23){\vector(1,-1){20}}
\put(-2,4){\vector(0,2){16}} \put(2,20){\vector(0,-2){16}}
\put(8,0){\vector(3,0){15}} \put(-8,0){\vector(-3,0){15}}

\put(0,25){\makebox(0,0){$\mathcal{L}$}}
\put(-27,0){\makebox(0,0){${\mathcal{L}_{1}}$}}
\put(27,0){\makebox(0,0){${\mathcal{L}_{2}}$}}
\put(0,0){\makebox(0,0){${\mathcal{L}_{1}\times\mathcal{L}_{2}}$}}
\put(-1,11){\makebox(-10,0){$\psi$}}
\put(-1,11){\makebox(13,0){$\tau$}}
\put(-15,16){\makebox(0,0){$\tau_{1}$}}
\put(15,16){\makebox(0,0){$\tau_{2}$}}
\put(-15,2){\makebox(0,0){$\pi_{1}$}}
\put(15,2){\makebox(0,0){$\pi_{2}$}}
\end{picture}
\caption{The different maps between $\mathcal{L}_{1}$,
$\mathcal{L}_{2}$, ${\mathcal{L}_{1}\times\mathcal{L}_{2}}$, and
$\mathcal{L}$. $\pi_{1}$ and $\pi_{2}$ are the canonical
projections.}\label{f:maps}
\end{center}

\end{figure}

The lack of injectivity of the $\tau_{i}$ may be physically
recognized from the fact that the state of the whole system encodes
information about correlations between its parts. It is again useful
to make a comparison with the classical case in order to illustrate
what is happening. The same as in classical mechanics, we have atoms
in $\mathcal{L}$ which are tensor products of atoms of
$\mathcal{L}_{1}$ and $\mathcal{L}_{2}$. But in contrast to
classical mechanics, entangled states originate atoms of
$\mathcal{L}$ which cannot be expressed in such a way and thus,
loosely using a topological language, we may say the fiber of the
projection $\tau_{i}$ is much bigger than that of its classical
counterpart.

It is important to note that the projection function $\tau$ cannot
be properly defined  within the frame of the traditional approach of
QL because there was no place for improper mixtures in
$\mathcal{L}_{vN}$, where they have to be defined as functions over
the sublattices. On the contrary, mixtures are \emph{elements} of
the lattices $\mathcal{L}$ and $\mathcal{L}_{i}$, and thus we can
define the projections from the lattice of the whole system to the
lattices of the subsystems mapping the states of $S$ into the
corresponding states of $S_{i}$. This enables a more natural
approach when compound systems are considered from a quantum logical
point of view.

It is interesting also to analyze the functions $\Psi\circ\tau$ and
$\tau\circ\Psi$.

\begin{prop}
$\tau\circ\Psi=Id$.
\end{prop}

\begin{proof}
Let us see it in terms of good representatives:
$$\tau_1(\Psi([S],[T]))=\tau_1([S\otimes T])=[\mbox{tr}_1(S\otimes T)]=[S]$$
$$\tau_2(\Psi([S],[T]))=\tau_2([S\otimes T])=[\mbox{tr}_2(S\otimes T)]=[T]$$
Then $\tau(\Psi([S],[T]))=([S],[T])$.
\end{proof}

It is clear, from a physical point of view, that $\Psi\circ\tau$ is
not the identity function: when we take partial traces information
is lost that cannot be recovered by making products of states. This
can be summarized as \emph{``going down and then going up is not the
same as going up and then going down''} (another way to express
quantum non-separability). We show these maps in Figure
\ref{f:maps}.

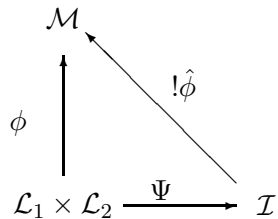
\begin{figure}\label{f:maps2}
\begin{center}
\unitlength=1mm
\begin{picture}(15,3)(0,0)
\put(23,3){\vector(-1,1){20}} \put(0,4){\vector(0,2){16}}
\put(8,0){\vector(3,0){15}}

\put(0,25){\makebox(0,0){$\mathcal{M}$}}
\put(27,0){\makebox(0,0){$\mathcal{I}$}}
\put(0,0){\makebox(0,0){${\mathcal{L}_{1}\times\mathcal{L}_{2}}$}}
\put(-1,11){\makebox(-10,0){$\phi$}}
\put(16,16){\makebox(0,0){$!\hat{\phi}$}}
\put(13,2){\makebox(0,0){$\Psi$}}
\end{picture}
\caption{This is a commutative diagram. $(\mathcal{I},\Psi)$ is the
lattice tensor product (in categorical terms) of $\mathcal{L}_1$ and
$\mathcal{L}_2$.}
\end{center}

\end{figure}

Let us finally make an observation about the image of $\Psi$.
Consider the category of lattices as objects and lattice-morphisms
as arrows. A bi-morphism is a morphism in each variable, and
proposition \ref{bimorfism} ensures that $\Psi$ is a bi-morphism.
Let us define $\mathcal{I}$ as the lattice generated by $Im(\Psi)$
inside $\mathcal{L}$. Then, the following relationship holds between
$\mathcal{I}$, $\mathcal{L}_{1}$ and $\mathcal{L}_{2}$ (See Figure
\ref{f:maps2}):

\begin{prop}
$(\mathcal{I},\Psi)$ is the lattice tensor product (in categorical
terms) of $\mathcal{L}_1$ and $\mathcal{L}_2$. That is, it satisfies
the following universal property: for every bi-morphism of lattices
$\phi:\mathcal{L}_1\times\mathcal{L}_2\rightarrow \mathcal{M}$ there
exists a unique $\hat{\phi}:\mathcal{I}\rightarrow \mathcal{M}$ such
that $\hat{\phi}\Psi=\phi$. And if $(\mathcal{I}',\Psi')$ is another
product then they are isomorphic by a unique isomorphism.
\end{prop}
\begin{proof}
Let $\phi:\mathcal{L}_1\times\mathcal{L}_2\rightarrow \mathcal{M}$ a
bi-morphism where $\mathcal{M}$ is an arbitrary lattice. Given that
$Im(\Psi)$ lattice-generates $\mathcal{I}$ we can define
$\hat{\phi}$ over the elements of the form $[S\otimes T]$:
$$\hat{\phi}([S\otimes T]):=\phi([S],[T])$$
Note that it is unique by definition and $\hat\phi\Psi=\phi$.

The unicity of $(\mathcal{I},\Psi)$ follows from a standard
categorical argument: Given that $\Psi'$ is a bi-morphism we have
$\hat{\Psi'}\Psi=\Psi'$ because $\Psi$ has the universal property.
Given that $\Psi'$ also has the universal property we have
$Id_{\mathcal{I}'}\Psi'=\Psi'$. The same holds for $\Psi$, that is
$\hat{\Psi}\Psi'=\Psi$ and $Id_{\mathcal{I}}\Psi=\Psi$. Note that
$\hat{\Psi'},\hat{\Psi},Id_{\mathcal{I}'},Id_{\mathcal{I}}$ are all
unique having this property. Given that
$\hat{\Psi'}\hat{\Psi}\Psi'=\Psi'$ and
$\hat{\Psi}\hat{\Psi'}\Psi=\Psi$ then we have:

$$\hat{\Psi}\hat{\Psi'}=Id_{\mathcal{I}}\quad
\hat{\Psi'}\hat{\Psi}=Id_{\mathcal{I}'}$$

\noindent So $\mathcal{I}$ and $\mathcal{I}'$ are isomorphic by a
unique isomorphism.
\end{proof}

\section{Conclusions}\label{s:Conclusions}

In this article we have shown that it is possible to construct a
lattice theoretical framework which incorporates improper mixtures
as atoms. This is done in order to overcome a problem of the
standard QL formalism posed in Section \ref{s:explicamosporque},
namely that the conjunction of all actual properties of the system
does not yield the actual state of the system when compound systems
are considered. We showed that this is directly linked with the fact
that QL treats improper mixtures as measures over the projection
lattice, in an analogous way as classical statistical distributions
are measures over the phase space. But alike classical mixtures,
improper mixtures in quantum mechanics do not admit an ignorance
interpretation, and this was at the origin of the problems posed in
Section \ref{s:explicamosporque}. Our construction is a quite
natural extension of the von Neumann lattice, and its properties and
characteristics are consistent with the constraints imposed by
quantum mechanics. More precisely, in the standard quantum logical
approach, when the whole system is in a pure entangled state there
are no elements available in the lattices of the subsystems to
represent the states of the subsystems as elements of the lattice.
This is expressed in the absence of projection functions which map
the states of the lattice of the whole system to the the states of
the lattices of the subsystems which satisfy in turn, to be
compatible with the physical description. Alike the standard
approach, the projections defined in the frame of the enlarged
structure satisfy this condition. They are also the canonical ones
in the sense that they are constructed using partial traces, in
accordance with the quantum formalism. This was shown in Section
\ref{s:projections}.

Traditionally, the difference between classical and quantum lattices
is said to be that the classical lattice is a Boolean lattice  while
von Neumann lattice is an orthomodular one. We claim that this is
not the only difference, the other one --although not independent--
being their behavior with respect to the coupling of two or more
systems. The necessity of the enlargement of the von Neumann lattice
in order to preserve the condition that the meet of actual
properties defines the state of the system may be seen as an
algebraic expression of the existence of entanglement. The approach
presented here shows, in an algebraic fashion, the radical
difference between quantum mechanics and classical mechanics when
two systems interact. If the systems are classical, no non-trivial
enlargement of the lattice is needed even in the presence of
interactions. It is enough in order to describe all relevant physics
about the subsystems. But the existence of entanglement in quantum
mechanics forces an enlargement of the state space of pure states to
the convex set $\mathcal{C}$ to deal with the states of subsystems
and thus the enlargement of $\mathcal{L}_{vN}$. A possible candidate
to fulfill this task, namely the lattice $\mathcal{L}$, has been
presented in this work and the relations among $\mathcal{L}$ and
$\mathcal{L}_i$ have been analyzed.  We think that paying more
attention to this kind of approaches would shed new light on the
algebraic properties of quantum non-separability.

\vskip1truecm

\noindent {\bf Acknowledgements} \noindent This work was partially
supported by the grant PIP N$^o$ 6461/05 (CONICET).

\end{document}